\documentclass{jfm}

\usepackage{microtype}
\usepackage{amsmath}
\usepackage{pifont}
\usepackage{natbib}

\usepackage{epsfig}
\usepackage{graphicx}

\usepackage{array}

\newcommand*{\Ca}{\text{Ca}}
\newcommand*{\Camac}{\Ca_\text{macro}} 
\newcommand*{\Cacl}{\Ca_\text{cl}}

\newcommand*{\eg}{\emph{e.g. }}

\newcommand{\de}{\textrm{d}}
\newcommand{\dd}[2]{\frac{\de#1}{\de#2}}
\newcommand{\pd}[2]{\frac{\partial#1}{\partial#2}}
\newcommand{\Oh}{\mathcal{O}}
\newcommand{\thetaapp}{\theta_\mathrm{app}}

\usepackage[usenames, dvipsnames]{xcolor}

\title[Distinguished limits in moving contact lines]{On the distinguished limits of the Navier slip model of  the moving contact line problem}

\author[Ren, Trinh, and E]%
{W\ls E\ls I\ls Q\ls I\ls N\ls G\ns R\ls E\ls N$^{1, \,2}$,\ns
P\ls H\ls I\ls L\ls I\ls P\ls P\ls E\ns H.\ns T\ls R\ls I\ls N\ls H$^3$ 
\and W\ls E\ls I\ls N\ls A\ls N\ns E$^{4, \,5}$}

\affiliation{%
$^1$ Department of Mathematics, National University of Singapore, Singapore 119076 \\[\affilskip]
$^2$ Institute of High Performance Computing, A*STAR, Singapore 138632 \\[\affilskip]
$^3$ Oxford Centre for Industrial and Applied Mathematics, Mathematical Institute, \\ University of Oxford, Oxford OX2 6GG, UK  \\[\affilskip]
$^4$ Department of Mathematics and Program in Applied and Computational Mathematics \\ Princeton University, Princeton 08544-1000, USA \\[\affilskip]
$^5$ School of Mathematics, Peking University, Beijing 100871, P.R. China}

\pubyear{XXXX}
\volume{YYY}
\pagerange{xxx--xxx}
\date{\today}

\begin{document}

\maketitle
\begin{abstract}
When a droplet spreads on a solid substrate, it is unclear what are the correct boundary conditions to impose at the moving contact line. The classical no-slip condition is generally acknowledged to lead to a non-integrable singularity at the moving contact line, for which a slip condition, associated with a small slip parameter, $\lambda$, serves to alleviate. In this paper, we discuss what occurs as the slip parameter, $\lambda$, tends to zero. In particular, we explain how the zero-slip limit should be discussed in consideration of two distinguished limits: one where time is held constant $t = \Oh(1)$, and one where time tends to infinity at the rate $t = \Oh(|\log \lambda|)$. The crucial result is that in the case where time is held constant, the $\lambda \to 0$ limit converges to the slip-free equation, and contact line slippage occurs as a regular perturbative effect. However, if $\lambda \to 0$ and $t \to \infty$, then contact line slippage is a leading-order singular effect.
\end{abstract}



\section{Introduction}

\noindent The moving contact line problem is explained as follows: the theory of traditional macroscopic fluid mechanics imposes the requirement that the velocity of a fluid in contact with a solid substrate must be equal to the velocity of the substrate (the `no slip condition'). However, this condition is obviously violated at a moving contact line, such as what occurs for a spreading droplet. In order to resolve this difficulty, the no-slip condition can be changed to an alternative condition that allows for slip. The challenge in resolving the moving contact line problem is to: (i) better understand the current slip models, their advantages and disadvantages; and (ii) propose alternative slip models that better represent the physics. In this paper, we shall focus on the former problem, and in particular, we discuss the distinguished nature of the zero-slip limit.

Here, we shall deal exclusively with the case that the contact line dynamics are modeled using the classic Navier slip condition. In two dimensions, where $u$ is the velocity parallel to the plane surface, and $z$ is measured normally away from it, this condition imposes 
\begin{equation} \label{navier}
u = \lambda \pd{u}{z},
\end{equation}

\noindent for a fluid in contact with a solid boundary at rest, and $\lambda$ is the slip coefficient, which is a measure of the length over which slip is significant. There are a multitude of papers in the literature on the asymptotics of the contact line problem as the slip parameter tends to zero (see \emph{e.g.} \cite{voinov_1976}, \cite{hocking_1982}, and \cite{lacey_1982}), and our paper seeks to highlight the idea of the non-uniformity of the perturbation methods as slip tends to zero and for different choices of time scales. This is most similar to the study of \cite{king_2001}, and \cite{flitton_2004}. We provide a more comprehensive listing of the vast literature on the topic of the moving contact line in \S{\ref{literature}}.

More specifically, in this work, we wish to demonstrate that when time is held at $\Oh(1)$, the dynamics of contact line spreading converges to a slipless equation as $\lambda \to 0$. However, in the limit $t \to \infty$, a re-scaling of time is necessary. Thus, convergence \emph{can} be achieved in the zero slip limit by using a slip-free equation, but only at finite time.  Our analysis seeks to explore this idea of non-uniformity using a combination of asymptotic techniques, and also accurate numerical results which clearly show the expected limiting behaviours in the singular regime. 

We present an asymptotic analysis of the lubrication equations for a droplet spreading under the effect of surface tension. In particular, there are two regimes: 
\begin{subequations}
\begin{align}
(i)& \text{ $\lambda \to 0$ and $t = t^*$ fixed} \\
(ii)& \text{ $\lambda \to 0$ and $t \gg 1$}
\end{align}
\end{subequations}
We find that in regime (i), contact line slippage is (almost) a `regular' perturbative effect---that is to say, as $\lambda \to 0$, the macroscopic motion of the droplet converges to the slip-less equation ($\lambda = 0$), and the apparent contact angle, $\theta_\text{app}$ converges to a value which can be determined solely from solving this particular equation. The apparent contact angle is not influenced by the microscopic conditions. In this regime, the contact line displacement tends to zero as slip tends to zero, and any contact line slippage is a higher-order effect within the macroscopic region. Note that slippage remains a leading-order contribution within the inner region near the contact line.

However, in the distinguished limit which involves the dual limit $\lambda \to 0$ and $t \to \infty$, the solutions in region (i) are no longer valid, and the asymptotic approximations in this regime become disordered. A re-scaling of time is necessary; once time is re-scaled, we recover the equivalent analyses of others [\emph{c.f.} \cite{hocking_1981sliding}, \cite{hocking_1983}, and \cite{coxrg_1986}], and contact line displacement becomes a significant effect. In particular, $\theta_\text{app}$ is now a function of the unknown contact line location and depends on the microscopic properties of the substrate. As is well known through these previous works, the contact line speed is logarithmically small in the slip number, and involves the key quantity
\begin{equation}
    \epsilon = 1/|\log \lambda|.
\end{equation}

\begin{figure}
\centering
\includegraphics{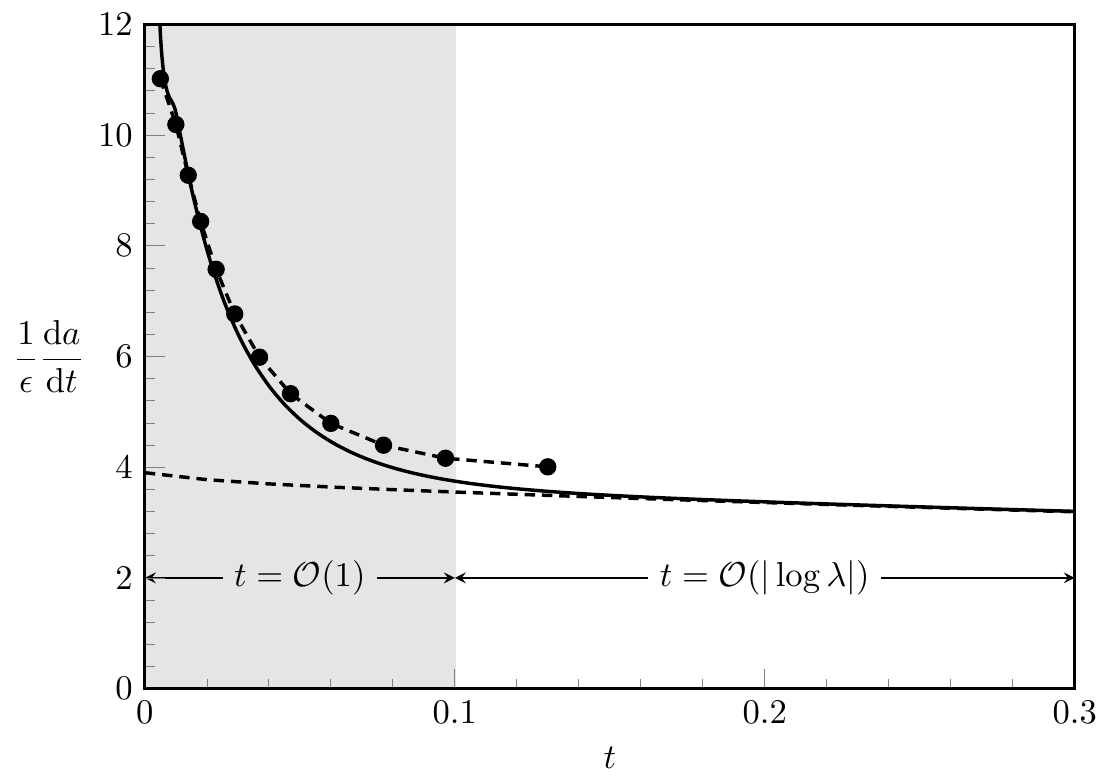}
\caption{(Solid) Re-scaled velocity, $\dot{a}(t)/\epsilon = \dot{a}(t) |\log \lambda|$ as a function of time for a spreading droplet released far from its quasi-static state. (Dashed) The classical quasi-static prediction of contact line speed via \eqref{ode_a} and \eqref{anglespeed_class} is only valid for large times (as the slip, $\lambda \to 0$). (Dashed markers) For $t = \Oh(1)$, the slip-free formulation using \eqref{anglespeed} and \eqref{anglespeed_final1} provides a better fit. Numerical computations are for the initial profile \eqref{h0} and $\lambda = 9 \times 10^{-7}$. The details of this image are discussed in \S\ref{hocking}. \label{bigpic}}
\end{figure}

The principle result of this paper is shown in Figure \ref{bigpic}, which plots a re-scaled contact line velocity, $\dot{a}(t)/\epsilon$, as a function of time for a spreading droplet. The graph demonstrates that the contact line moves rapidly initially, then slows as time increases. The two time scales determining the dynamics are clearly visible, and the asymptotic approximations developed in the paper are shown as determining the contact line movement in each respective region.

This notion of multiple scalings of time influencing the resultant contact line asymptotics allows us to better understand the nature of the zero slip limit. For example, we seek to better understand the early work by \cite{moriarty_1992}, who studied the quasi-static \cite{greenspan_1978} of the moving contact line, and sought to understand the relationship between the slip coefficient and the necessary finite difference grid-spacing to achieve convergent results. They explained that 
\begin{quote}\emph{\ldots converged finite results, if slip is ignored, can never be obtained. This is the numerical manifestation of the non-integrable force singularity at a moving contact-line when slip is not permitted.}\end{quote}

\noindent Thus one of the goals of this paper is to demonstrate that if time is held at $\Oh(1)$, then converged numerical results can in fact be obtained; a zero slip condition can be applied to the macroscopic model. 

We shall begin in \S{\ref{literature}} by briefly reviewing the literature behind theories on the moving contact line, with particular emphasis on the classic macroscopic models, molecular models, and mesoscopic models. The discussion in this paper will focus on the simplest case of a thin, spreading droplet, and the mathematical formulation is presented in \S{\ref{mathform}}. We analyze the $t = \Oh(1)$ problem in \S{\ref{outer}} to \ref{matching}, and relate this to the classical analyses of, for example, workers such as \cite{hocking_1983} in \S{\ref{hocking}}. We conclude with a discussion in \S\ref{discuss}, focusing on the topic of the role of distinguished limits in more complicated systems involving contact lines. 


\subsection{A variety of contact line models} \label{literature}

\noindent It would be misleading for us to proceed without fully acknowledging the great body of literature that already exists on the moving contact line problem. Theoretical models of moving contact lines can be roughly divided into three categories: (i) molecular kinetic models, (ii) molecular dynamic models, and (iii) hydrodynamic models. A sampling of reference works, separated by these three classifications, is given in Table \ref{tab:res}. 

The molecular kinetics model of contact lines was first proposed by \cite{blake_1969}, and later extended by \cite{blake_1993} and \cite{blake_2002}. In this model, the dynamics of the contact line is described by an absorption and desorption process of the fluid molecules on the solid surface. The theory provides a quantitative description for the contact line friction at the microscopic scale, and gives a link between microscopic quantities, such as the frequency and length of molecular displacements, with the macroscopic behaviour of the dynamic contact angle. 

The search to better understand microscopic details of contact lines leads naturally to the idea of using molecular dynamics, and studies in this vein include the works by \cite{koplik_1988, koplik_1989}, \cite{thompson_1989}, \cite{blake_1997, blake_1999}, \cite{ren_2007} and \cite{coninck_2008}. The approach has been very successful and computations have revealed much in regards to the physical processes near the contact line. The disadvantage, however, is that such simulations are limited to systems of small scale and within small temporal intervals. As such, it remains difficult to relate molecular dynamics to the macroscopic scale.

Lastly, moving contact lines can be studied using hydrodynamic models, and this includes the classical works of, for example, \cite{huh_1971}, \cite{dussan_1974}, \cite{voinov_1976}, \cite{hocking_1982}, and \cite{coxrg_1986}, and the recent work \cite{ren_2010cont}. Such models will impose slip through a boundary condition on the macroscopic variables, and thus assumes the specification of an effective condition for the underlying microscopic mechanisms [\eg the Navier slip condition of \eqref{navier}]. The primary advantage of such approaches is that the usual governing equations (\eg Navier-Stokes) is used with little modification, except for a replacement of the no-slip condition. 

As a middle-ground between the molecular and classical macroscopic approaches, it is also possible to incorporate intermolecular forces and more detailed physics of the finite-width fluid interface into the hydrodynamics. Such mesoscopic continuum models include the diffuse interface models studied in \cite{jacqmin_2000}, \cite{pismen_2002}, \cite{Qian_2003}, and \cite{yue_2010}, as well as the work by \cite{shik_1997, shik_book} and \cite{billingham_2008}, where interface creation and destruction processes are modeled. 

\begin{table}
\begin{center}
\begin{tabular}{>{\centering\arraybackslash}p{3.3cm}cp{9cm}}
Kinetic theory && \cite{blake_1969}, \cite{blake_1993} \cite{blake_2002} \\[0.5\baselineskip]
Molecular dynamic && 
\cite{koplik_1988, koplik_1989}, 
\cite{thompson_1989}, 
\cite{blake_1997, blake_1999}, 
\cite{ren_2007},
\cite{coninck_2008}
 \\[0.5\baselineskip]
Hydrodynamic (macroscopic) && 
\cite{huh_1971}, 
\cite{dussan_1974}, 
\cite{voinov_1976},
\cite{greenspan_1978},
\cite{lacey_1982}, 
\cite{hocking_1982}, 
 \cite{hocking_1983, hocking_1992},
\cite{coxrg_1986}, 
\cite{haley_1991}, 
\cite{bertozzi_1994},
\cite{king_2001},
\cite{eggers_2004toward, eggers_2005existence}, 
\cite{eggers_2004characteristic},
\cite{ren_2010cont}  \\[0.5\baselineskip]
Hydrodynamic \mbox{(mesoscopic)} && 
\cite{shik_1997, shik_book},
\cite{anderson_1998},
\cite{jacqmin_2000}, 
\cite{pismen_2002}, 
\cite{Qian_2003},
\cite{wilson_2006}, 
\cite{billingham_2008},
\cite{yue_2010}\\[0.5\baselineskip]
\end{tabular}
\end{center}
\caption{A sample of works using the kinetic, molecular dynamic, or hydrodynamic theories of contact line motion. \label{tab:res} }
\end{table}

Our work in this paper is primarily inspired by the body of work following from \cite{lacey_1982} and \cite{hocking_1983}, and thus we shall focus on the standard classical hydrodynamic model with Navier slip. However, it is important that we mention that all three models of contact line motion are appreciated, and it is still an active area of research to establish the advantages and disadvantages of each of the models. For more details, see the reviews by \cite{dussan_1974}, \cite{pomeau_2002}, \cite{kistler_1993}, \cite{blake_2006}, \cite{lauga_2007}, as well as the collected volume edited by \cite{velarde_2011aa}.

\section{Mathematical Formulation} \label{mathform}

\noindent We shall consider the symmetrical spreading of a thin viscous droplet of height $z = h(x,t)$, over a flat surface, where the slip on the surface is governed by the Navier slip law \eqref{navier}. The governing equations (see \eg \cite{lacey_1982}) are given by
\begin{equation}
\frac{\partial h}{\partial t} + \frac{\partial}{\partial x} 
\left(h^2 \left( \frac{h}{3} + \lambda \right) 
\frac{\partial^3 h}{\partial x^3} \right) = 0,
\end{equation}
on the domain $0 \leq x \leq a(t)$. The droplet begins from an initial state $h(x, 0) = g(x)$, and is subject to symmetry boundary conditions at the origin,
\begin{equation}
\partial h/\partial x = 0 = \partial^3 h /\partial x^3 \quad \text{at} \quad x = 0 \label{cond1}. 	
\end{equation}
The height of the droplet vanishes at the moving edge
\begin{subequations} \label{contcond}
\begin{equation}
h = 0, \quad \text{at}\ x = a(t). \label{cond_0}	
\end{equation}
We assume that the equilibrium angle, $\theta_y \neq 0$ (partial wetting), and also, that the contact line $a(t)$ is advected according to the constitutive relation:
\begin{equation} \label{eq:CLcond}
\beta \dot{a} = \frac{1}{2} \Bigl[ (\partial h/\partial x)^2 
 - \theta_y^2 \Bigr] \qquad \text{at}\ x = a(t),
\end{equation}
\end{subequations}
where $\dot a =\de{a}/\de{t}$ is the velocity of the contact line. This constitutive law can be viewed as a force balance at the moving contact line, where the friction force on the left-hand side is balanced by the unbalanced Young stress on the right-hand side \citep{ren_2010cont}. Other constitutive laws for the advective behaviour are possible (see discussions in, for example, \citealt{haley_1991}), but the details of our analysis will be largely independent of this choice. The case of complete wetting is addressed in Appendix \ref{perfectwetting}. 

For convenience, we rescale the variables as follows: $\widehat{h} = 3h$, $\widehat{x} = 3 x$, $\widehat{a} = 3a$, and $\widehat{t} = t$. Writing $\widehat{\lambda} =9 \lambda$ and $\widehat{\beta} = \beta/3$, 
this has the effect of changing the equation to (dropping hats):
\begin{equation}  \label{eq:heq}
\frac{\partial h}{\partial t} + \frac{\partial }{\partial x} 
\left( h^2 \left( h + \lambda \right) \frac{\partial^3 h}{\partial x^3} \right) = 0,
\end{equation}
with boundary conditions \eqref{cond1} and \eqref{cond_0}, and the condition for the contact line \eqref{eq:CLcond}. Finally, we introduce local coordinates relative to the contact line. Letting $x = a(t) - X$ and $h(x, t) = H(X,t)$, the governing equation yield
\begin{equation} \label{govX}
\frac{\partial H}{\partial t} + \dot a \frac{\partial H}{\partial X} 
+ \frac{\partial}{\partial X}\left( H^2 
\left( H + \lambda \right) \frac{\partial^3 H}{\partial X^3} \right) = 0.
\end{equation}


\section{Asymptotic analysis of the outer region at $t = \Oh(1)$} \label{outer}

\noindent We are interested in the solution in the $\lambda \to 0$ limit; in this limit, the contact line speed tends to zero, so we make the expansion:
\begin{equation} \label{aseries}
a(t) = a_0 + \epsilon a_1(t) + \epsilon^2 a_2(t) + \ldots,
\end{equation}
where $a_0$ is the initial contact line location. 
We claim, and this can be verified \emph{a posteriori}, that $\epsilon \gg \lambda$. Thus, we expand $H = H_0 + \epsilon H_1 + \Oh(\epsilon^2, \lambda)$, where the first correction term is indeed $\Oh(\epsilon)$ with the assumption $\epsilon \gg \lambda$. Temporarily keeping the $\lambda$ term, we have at leading order
\begin{equation}
\frac{\partial H_0}{\partial t} + \frac{\partial}{\partial X}
\left( H_0^2 \left( H_0 + \lambda \right) \frac{\partial^3 H_0}{\partial X^3} \right) 
= 0.
\end{equation}

\subsection{Leading-order outer equation} \label{leadout_t1}

\begin{figure}
\centering
\includegraphics{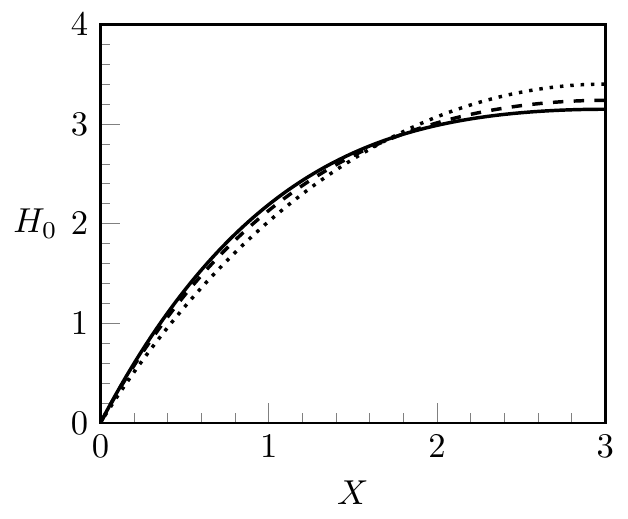} \\
\begin{minipage}{0.49\textwidth}
\includegraphics{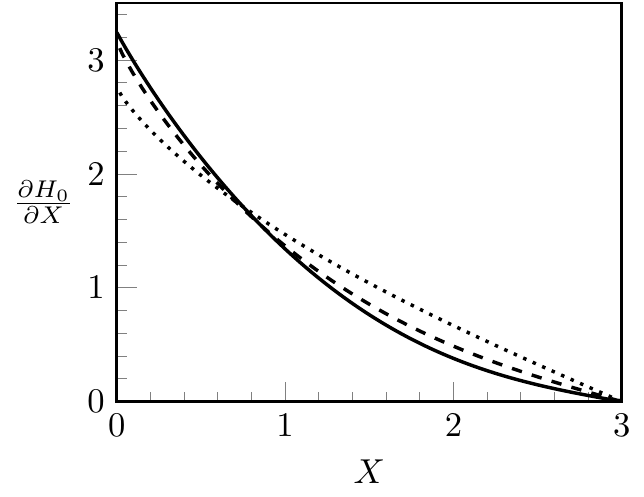}
\end{minipage}
\begin{minipage}{0.49\textwidth}
\vspace*{0pt}
\includegraphics{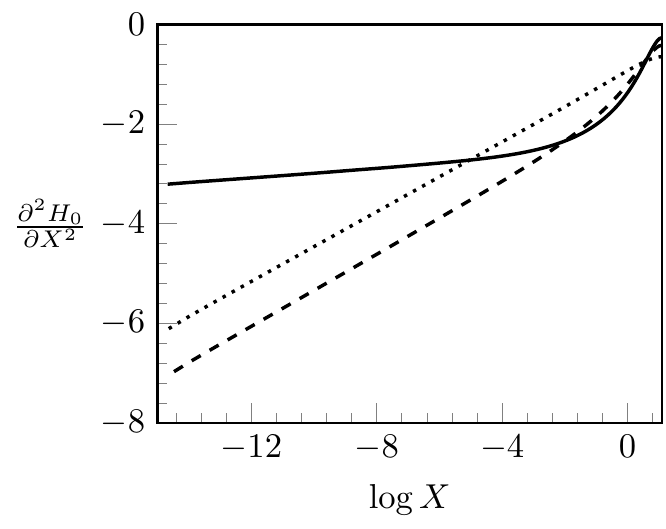}
\end{minipage}
\caption{(Upper panel) The solution to the leading-order outer equation \eqref{H0eqn} at 
$t=0.005$ (solid line), $t=0.01$ (dashed line) and $t=0.03$ (dotted line). 
The contact line is at $X=0$. (Bottom left) The first derivative of the solution versus $X$.
(Bottom right) The second derivative of the solution versus $\log X$. The initial condition used is \eqref{h0}. 
\label{fig:H0}}
\end{figure}

\noindent Away from $X = 0$, we may ignore the $\lambda$ slip term, and this gives for the outer approximation, 
\begin{equation} \label{H0eqn}
\frac{\partial H_0}{\partial t} + \frac{\partial}{\partial X}\left(
H_0^3 \frac{\partial^3 H_0}{\partial X^3} \right) = 0.
\end{equation}
One may solve (\ref{H0eqn}) using only the single contact line condition $H_0(0, t) = 0$. This would be consistent with the idea that the microscopic contact angle cannot be applied within this outer region. 
The numerical solution to equation \eqref{H0eqn}, and its first and second
spatial derivatives are shown in Figure \ref{fig:H0}.
Note that the slope remains well behaved as $X \to 0$, so 
$H_0$ provides a well-defined apparent contact angle (middle panel), given by
\begin{equation}
\thetaapp(t) \sim \frac{\partial H_0}{\partial X}\biggr\rvert_{X = 0}.
\end{equation}
Also the second derivative of the solution, \emph{i.e.} the curvature of the interface, diverges as $\log X$ as $X\rightarrow 0$, and this can be seen in the lower panel of Figure \ref{fig:H0}.

Based on the observation from the numerics, 
we make the series expansion in the limit that $X\to 0$:
\begin{equation} \label{H0ansatz}
H_0(X,t) = B_{10}(t)X + \sum_{i=2}^\infty \Bigl( B_{i0}(t) + B_{i1}(t) \log X \Bigr) X^i.
\end{equation}

\noindent From \eqref{H0eqn}, this gives for the first two orders:
\begin{subequations}
\begin{alignat}{3}
\Oh(X): \qquad && 4 B_{10}^3 B_{21} + \dot{B}_{10} &= 0, \label{BX} \\
\Oh(X^2 \log X): \qquad && 18 B_{10}^2 B_{21}^2 + 18 B_{10}^3 B_{31} 
  + \dot{B}_{21} &= 0,
\label{H0first}
\end{alignat}
\end{subequations}
where we used dots to denote the time derivative. Also, (\ref{BX}) gives the leading-order relation, $2 \thetaapp^3 (\partial^2H/\partial X^2)\log X + \de\thetaapp/\de{t} = 0$, between the divergent curvature with the apparent contact angle and its time evolution. 

In this paper, we use a semi-implicit finite difference scheme to numerically solve the partial differential equation \eqref{eq:heq} and its slip-free reduction \eqref{H0eqn}. Within this scheme, the spatial derivatives are treated implicitly, and the nonlinear terms explicitly. The numerical verification of the results in this paper presents a challenging problem (\emph{c.f.} further discussion of the issues in \cite{moriarty_1992}), and we use a refined mesh near the contact line to ensure convergent results. The scheme is detailed in Appendix \ref{sec:num}. The initial condition is mostly unimportant (so long as it begins away from the quasi-static state), and throughout this work, we shall use an equilbrium contact angle, $\theta_y = 1$, and initial condition
\begin{equation} \label{h0}
h(x,0) = 3 \cos(\pi x^2/18).
\end{equation}

\begin{figure} \centering
\includegraphics{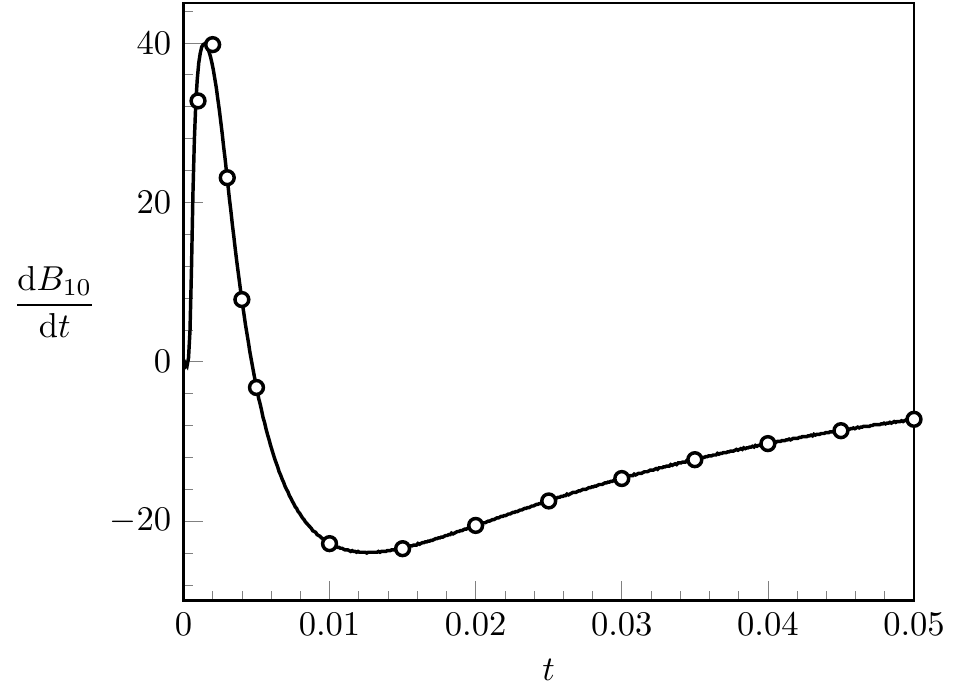}
\caption{Verification of the relation \eqref{BX} 
between the contact angle and the rate of divergence of the curvature.
The solid curve is $\dot{B}_{10}$, and the 
circles are $-4 B_{10}^3 B_{21}$, where $B_{10}$ and $B_{21}$ are computed
from the numerical solution $H_0$. \label{fig:B10t}}
\end{figure}

We now seek to verify the relation between the divergent curvature and the apparent angle in \eqref{BX}. The slip-free equation \eqref{H0eqn} is solved with the single contact line condition, $H_0(0, t) = 0$, and the profiles of $H_0$ and its derivatives are shown in Figure \ref{fig:H0}. At the origin, $X = 0$, by comparing $\partial H_0/\partial X$ with $X$, the value of $B_{10}$ is extracted, and from comparing $\partial^2 H_0/\partial X^2$ with $\log X$, the value of $B_{21}$ is extracted (the lower panel of Figure \ref{fig:H0}). Once the time-dependent  $B_{10}$ and $B_{21}$ have been computed, we can verify the relation \eqref{BX}. It is seen in Figure \ref{fig:B10t} that the numerical solution obeys the relation \eqref{BX} very well.

\subsection{First-order outer equation}

\noindent Turning to the next order, if we ignore the terms with $\lambda$, 
then we have
\begin{equation} \label{H1eqn}
\frac{\partial H_1}{\partial t} 
+ \dot{a}_1 \frac{\partial H_0}{\partial X}
+ \frac{\partial}{\partial X}
\biggl( H_0^3 \frac{\partial^3 H_1}{\partial X^3} + 
3H_0^2 H_1 \frac{\partial^3 H_0}{\partial X^3}\biggr) = 0.
\end{equation}
We shall assume that $\dot{a}_1(t) = \Oh(1)$, and the boundary condition also requires that $H_1(0, t) = 0$. The only consistent leading-order balance of the four groups of terms in \eqref{H1eqn} occurs between the second and forth terms. In this case, we may verify that as $X \to 0$, the expansion for $H_1$ follows $H_1 = \Oh(X \log X)$. The correct expansion for $H_1$ as $X \to 0$ is
\begin{equation} \label{H1ansatz}
H_1(X,t) = C_{10}(t)X + C_{11}(t) X \log X + \sum_{i=2}^\infty 
\Bigl( C_{i0}(t) + C_{i1}(t) \log X \Bigr) X^i,
\end{equation}
where the functions $C_{ij}(t)$ are to be determined. This gives the two leading order equations: 
\begin{subequations}
\begin{alignat}{3}
\Oh(1): \qquad && \dot{a}_1 B_{10} - B_{10}^3 C_{11} &= 0, \\
\Oh(X \log X): \qquad && 2 \dot{a}_1 B_{21} + 6 B_{10}^2 B_{21} C_{11} 
    + \dot{C}_{11} &= 0.
\label{H1first}
\end{alignat}
\end{subequations}
The first equation allows us to solve for $C_{11}$, while the second one allows us to solve for $B_{21}$. In summary, combining equations (\ref{H0ansatz})--(\ref{H0first}) and 
(\ref{H1ansatz})--(\ref{H1first}), we have as $X \to 0$, 
the inner limit of the outer approximation:
\begin{multline} \label{Hexpouter}
H_\text{out $\to$ in} = \biggl\{ B_{10} X + B_{21} X^2 \log X + \ldots \biggr\} \\ 
+ \epsilon \left\{ \left( \frac{\dot{a}_1}{B_{10}^2(t)} \right) X \log X 
  + C_{10}X + \cdots \right \}.
\end{multline}

\section{Asymptotic analysis of the inner region at $t = \Oh(1)$} \label{inner}

\noindent For the outer approximation of the previous section, 
we did not apply the exact wall condition given by \eqref{eq:CLcond}. 
Moreover, it should be clear that the expression (\ref{Hexpouter}) 
breaks down when $\epsilon \log X = \Oh(1)$, or 
when $X = \Oh(e^{-1/\epsilon})$; in this smaller region, the terms in the outer approximation begin to re-arrange. However, when $H$ and $X$ are small, then there is an inner region whose size is determined by the slip parameter, $\lambda$. Thus the correct scaling for the contact line speed, $\epsilon$, is given by precisely balancing the size of the slip region with the predicted breakdown of the outer approximation, and we require $\lambda = \Oh(e^{-1/\epsilon})$. We thus set  
\begin{equation}
\epsilon = 1/|\log \lambda|.
\end{equation}
For the inner region, we re-scale $H = \lambda \overline{H}$ and $X = \lambda s$, then (\ref{govX}) gives
\begin{equation} \label{govXin}
\lambda \frac{\partial \overline{H}}{\partial t} + \dot{a} 
\frac{\partial \overline{H}}{\partial s} + \frac{\partial}{\partial s} 
\left[ \overline{H}^2 \left( \overline{H} + 1 \right) \frac{\partial^3 
\overline{H}}{\partial s^3} \right] = 0.
\end{equation}
We expand $\overline{H} = \overline{H}_0 + \epsilon \overline{H}_1 + \Oh(\epsilon^2, \lambda)$, and this gives the first two orders as
\begin{subequations}
\begin{gather}
\frac{\partial}{\partial s} \left[ \overline{H}_0^2 
\left( \overline{H}_0 + 1 \right) \frac{\partial^3 
\overline{H}_0}{\partial s^3} \right] = 0, \\
\dot{a}_1 \frac{\partial \overline{H}_0}{\partial s} 
+ \frac{\partial}{\partial s} \left[ \overline{H}_0^2 
\left( \overline{H}_0 + 1 \right) \frac{\partial^3 \overline{H}_1}{\partial s^3} + 
\left( 3\overline{H}_0^2 \overline{H}_1 + 2\overline{H}_0 \overline{H}_1 \right) 
\frac{\partial^3 \overline{H}_0}{\partial s^3}
\right] = 0.
\end{gather}
\end{subequations}
The necessary boundary conditions at $s = 0$ are given by (\ref{cond_0}) and \eqref{eq:CLcond}:
\begin{equation} \label{BCin}
\overline{H} = 0 \quad \text{and} \quad
\partial \overline{H}/ \partial s = \theta_y + 
  \epsilon \left( \beta \dot{a}_1/\theta_y \right) + \Oh(\epsilon^2).
\end{equation}
The leading-order problem is solved, giving
\begin{equation} \label{H0inner}
\overline{H}_0(s, t) = \theta_y s.
\end{equation}
The first-order problem can be integrated once and gives
\begin{equation} \label{Cteqn}
\dot{a}_1 \overline{H}_0 + \left(\overline{H}_0^3 + 
\overline{H}_0^2\right) \frac{\partial^3 \overline{H}_1}{\partial s^3} = C(t).
\end{equation}
With $\overline{H}_0$ given by (\ref{H0inner}), it can be verified \emph{a posteriori} that the third derivative 
of $\overline{H}_1$ is $\Oh(s^{-2})$ as $s\to 0$, so $C(t) \equiv 0$. The resultant equation is integrated for $\overline{H}_1$ and application of the boundary conditions \eqref{BCin} gives
\begin{multline}
\overline{H}_1(s,t) = C_1(t) s^2  + \dot{a}_1
\left( -\frac{s}{2\theta_y^2} + 
\frac{\beta s}{\theta_y} - \frac{s^2 \log s}{2\theta_y} \right. \\ 
\left. + \frac{\log(1 + \theta_y s)}{2\theta_y^3} + \frac{s\log(1 + \theta_y s)}{\theta_y^2} + \frac{s^2 \log(1+\theta_y s)}{2\theta_y} \right).
\end{multline}
We shall assume that $\overline{H}_1(s,t)$ does not diverge faster than $s \log s$ as $s \to \infty$ so we set
$C_1(t) = - \dot{a}_1 \log \theta_y/(2\theta_y)$, leaving us with the final first-order solution
\begin{multline} \label{H1inner}
\overline{H}_1(s,t) = \dot{a}_1 \left( -\frac{s}{2\theta_y^2} 
+ \frac{\beta s}{\theta_y} - \frac{s^2 \log \theta_y}{2\theta_y} 
 - \frac{s^2 \log s}{2\theta_y} \right. \\ 
\left. + \frac{\log(1 + \theta_y s)}{2\theta_y^3} + 
\frac{s\log(1 + \theta_y s)}{\theta_y^2} + 
\frac{s^2 \log(1+\theta_y s)}{2\theta_y} \right).
\end{multline}

\noindent Notice that as $s \to 0$, the third derivative of 
$\overline{H}_1$ is $\Oh(s^{-1})$, so the assumption made 
after (\ref{Cteqn}) is verified. All together, as $s \to \infty$, we have the outer limit of the inner solution:
\begin{equation} \label{inout}
\overline{H}_{\text{in $\to$ out}} \sim \theta_y s + 
\epsilon \dot{a}_1
 \left[ \left(\frac{1}{\theta_y^2}\right)s \log s + 
\left( \frac{\beta}{\theta_y} + \frac{\log \theta_y}{\theta_y^2} \right) s +\cdots \right].
\end{equation}

\section{Asymptotic analysis of the intermediate region at $t = \Oh(1)$}

\noindent In general, we cannot expect $H$ to match directly with $\overline{H}$ (since the out-to-in limit is a time-dependent angle, and the in-to-out limit is a specified, constant angle). We need an intermediate region to perform the matching, and this is given by the larger $\epsilon$ parameter. In this region, we write
\begin{equation} \label{scaling_interm}
s = e^{z/\epsilon}, \qquad \overline{H} = Q(z,t) e^{z/\epsilon}.
\end{equation}
where $0 < z < 1$ provides the intermediate scaling between inner and outer regions. We must now change (\ref{govXin}) to make use of differentiation in $z$. 
Before doing this, however, let us examine the first time-dependent term 
in (\ref{govXin}). This term becomes $\lambda \partial \overline{H}/\partial t = e^{(z - 1)/\epsilon}\partial Q/\partial t$. Within the intermediate region, this term is exponentially small, and should thus be ignored. Thus within the intermediate region, we have
\begin{equation} 
\dot{a} \frac{\partial \overline{H}}{\partial s} + \frac{\partial}{\partial s} 
\left( \overline{H}^2 \left( \overline{H} + 1 \right) 
\frac{\partial^3 \overline{H}}{\partial s^3} \right) = 0.
\end{equation}
%
Integrating the equation once and setting the constant of integration to zero, 
then re-writing in intermediate variables gives
\begin{equation}
\dot{a} + Q\left( Q + e^{-z/\epsilon}\right) 
\left( -\epsilon \frac{\partial Q}{\partial z} 
+ \epsilon^3 \frac{\partial^3 Q}{\partial z^3}\right) = 0.
\end{equation}
We ignore the exponentially small term and expand the velocity. This gives
\begin{equation}
\left( \epsilon \dot{a}_1 + \epsilon^2 \dot{a}_2 
+ \Oh(\epsilon^3) \right) 
+ Q^2\left(-\epsilon \frac{\partial Q}{\partial z} 
+ \epsilon^3 \frac{\partial^3 Q}{\partial z^3} \right) = 0.
\end{equation}
Notice that up to order $\epsilon^3$ in the above equation, we can derive a portion of the solution as $Q^3 = \left(c_0 + \epsilon c_1\right) + 3(\dot{a}_1  + \epsilon \dot{a}_2) z + \Oh(\epsilon^2)$, which gives
\begin{equation}
Q(z, t) = \left(c_0 + 3 \dot{a}_1 z\right)^{1/3} 
+ \epsilon \left( \frac{c_1 + 3 \dot{a}_2 z}{3(c_0 + 3 \dot{a}_1 z)^{2/3}} \right) 
+ \Oh(\epsilon^2).
\end{equation}
We can thus write the asymptotic expansion of the intermediate solution as 
\begin{equation} \label{interm}
\overline{H}_\text{interm} = \left(c_0 + 3 \dot{a}_1 z\right)^{1/3}s 
+ \epsilon \left( \frac{c_1 + 3 \dot{a}_2 z}{3(c_0 + 3 \dot{a}_1 z)^{2/3}} 
\right) s + \Oh(\epsilon^2).
\end{equation}

\section{Matching of inner, intermediate, and outer solutions} \label{matching}

\noindent In order to perform the matching between the solution in the intermediate region (\ref{interm}) and the solution in the inner region (\ref{inout}), we apply van Dyke's matching rule \citep{vandyke_book}: the two-term expansion of the intermediate solution (2:int), re-written in inner coordinates and re-expanded to two terms (2:inner), is equal to the two-term inner expansion, re-written in intermediate coordinates, and re-expanded to two terms. Or simply $\text{(2:inner)(2:int)} = \text{(2:int)(2:inner)}$. We thus have
\begin{align}
\text{(2:inner)(2:int)} &= 
\left[c_0 + \epsilon \left(3 \dot{a}_1 \log s\right)\right]^{1/3}s 
+ \epsilon 
\left[\frac{c_1 + 3 \dot{a}_2 \epsilon \log s}
{3(c_0 + 3 \dot{a}_1 \epsilon \log s)^{2/3}} \right]s   \notag \\
&= c_0^{1/3} s + \epsilon \left[ \left(\frac{\dot{a}_1}{c_0^{2/3}}\right)s \log s 
+ \left(\frac{c_1}{3 c_0^{2/3}}\right) s + \cdots \right].
\end{align}
which is matched to
\begin{align}
\text{(2:int)(2:inner)} &= \theta_y s + \epsilon \dot{a}_1
 \left[ \left(\frac{1}{\theta_y^2}\right)s \log s + 
\left( \frac{\beta}{\theta_y} + \frac{\log \theta_y}{\theta_y^2} \right) s+ \cdots \right],
\end{align}
and yields
\begin{equation} \label{c0c1}
c_0 = \theta_y^3 \qquad \text{and} \qquad
c_1 = 3\dot{a}_1\left(\beta \theta_y + \log \theta_y\right).
\end{equation}

This leaves the matching of intermediate and outer solutions. Substituting the outer variables $H = \lambda \overline{H}$ and $X = \lambda s$ into the intermediate solution (\ref{interm}), we have
\begin{equation} \label{Hinterm}
H_\text{interm} = \left[ c_0 + 3\dot{a}_1 \left(1 + \epsilon \log X\right) 
\right]^{1/3}X + 
\epsilon \left[ \frac{c_1 + 3\dot{a}_2 \left(1 + \epsilon \log X\right)}
{3 \left(c_0 + 3\dot{a}_1 \left(1 + \epsilon \log X\right)\right)^{2/3}} \right] X 
+ \cdots.
\end{equation}
The two-term intermediate limit (2:int), expressed in outer variables and re-expanded to two terms (2:out), gives
\begin{multline} \label{int2out}
\text{(2:out)(2:int)} = \left( c_0 + 3\dot{a}_1 \right)^{1/3} X \\ 
+ \epsilon 
\left[ \left(\frac{\dot{a}_1 }{(c_0 + 3 \dot{a}_1 )^{2/3}} \right) X \log X 
+ \left( \frac{c_1 + 3\dot{a}_2}{3(c_0 + 3\dot{a}_1 )^{2/3}}\right) X \right],
\end{multline}
\noindent whereas from (\ref{Hexpouter}), we have
\begin{equation}
\text{(2:int)(2:out)} = \left( B_{10} X + \cdots \right) 
+ \epsilon \left[ \left( \frac{\dot{a}_1}{B_{10}^2(t)} \right) X \log X 
 + C_{10}X + \cdots \right].
\end{equation}
Thus, we have the two equations
\begin{subequations}
\begin{gather}
B_{10}^3(t) - \theta_y^3 = 3\dd{a_1}{t},  \label{anglespeed} \\ 
B_{10}^2(t) \cdot C_{10}(t) = \dot{a}_1 \left(\beta \theta_y + 
\log \theta_y \right) + \dot{a}_2. \label{anglespeed_more}
\end{gather}
\end{subequations}

\begin{figure} \centering
\includegraphics{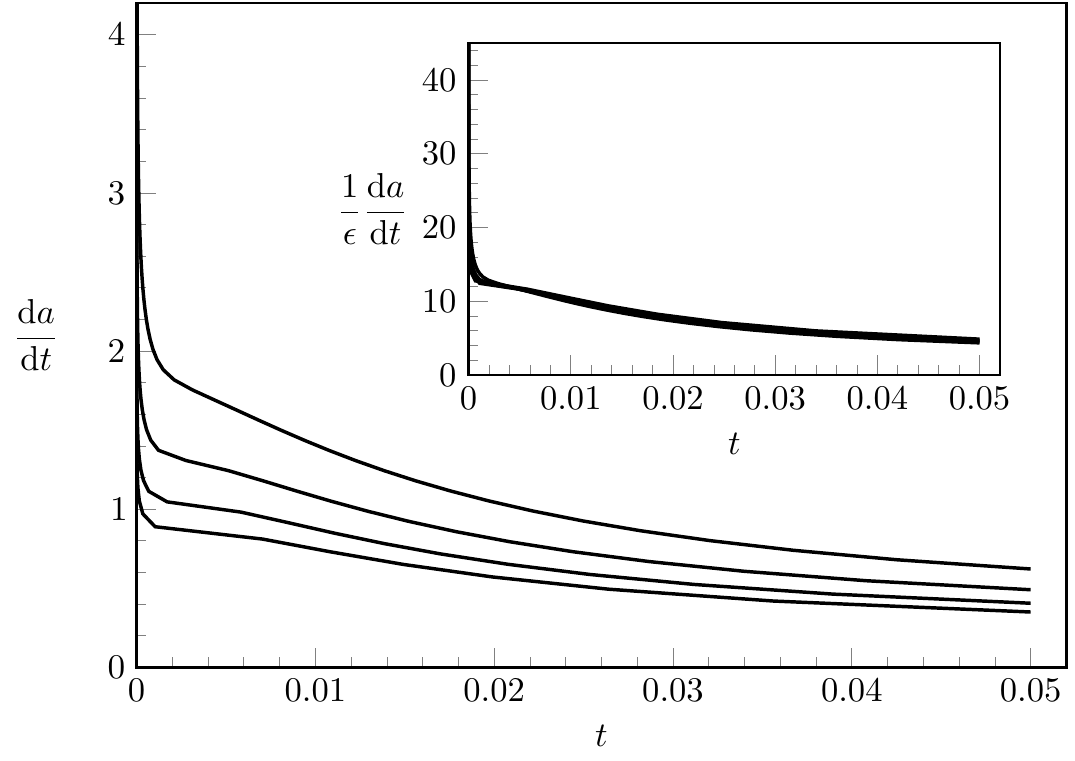} \\
\includegraphics{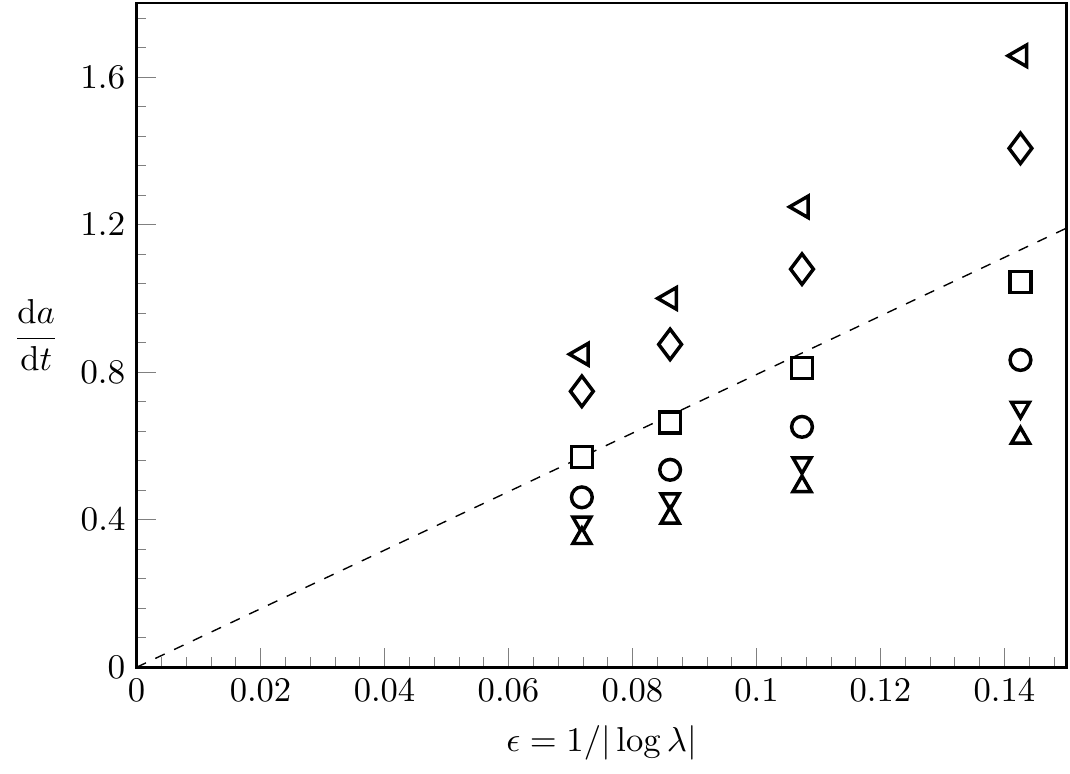}
\caption{(Upper panel) The velocity of the contact line versus time.
From top to bottom, the four curves are the velocity for $\lambda = 9\times 10^{-4}, 9\times 10^{-5}, 9\times 10^{-6}$ and $9\times 10^{-7}$, respectively. (Lower panel) The velocity of the contact line versus  $\epsilon = 1/|\log\lambda|$, at the time $t=0.005$ (left triangles), 0.01 (diamonds), 0.02 (squares), 0.03 (circles), 0.04 (down triangles), and 0.05 (up triangles). \label{fig:vel}}
\end{figure}

The relation between the contact angle and the contact line speed, \eqref{anglespeed}, is verified by numerics. The left-hand side of the relation follows from the computation of the leading-order slip-free outer solution $H_0$ in \S\ref{outer}. The right-hand side requires an accurate extraction of the limiting contact line velocity as $\lambda \to 0$. In order to obtain this value, we plot in Figure \ref{fig:vel} (lower panel) the velocity at fixed values of time and in decreasing values of the slip length. Note that the plot of the velocity versus $\epsilon = 1/|\log\lambda|$ appears to tend to a straight line passing through the origin. The right-hand side of \eqref{anglespeed}, $\dot{a}_1$, is estimated as the slope of the line joining the origin and the last data point ($\lambda = 9\times 10^{-7}$) at the different times.

\begin{figure} \centering
\includegraphics{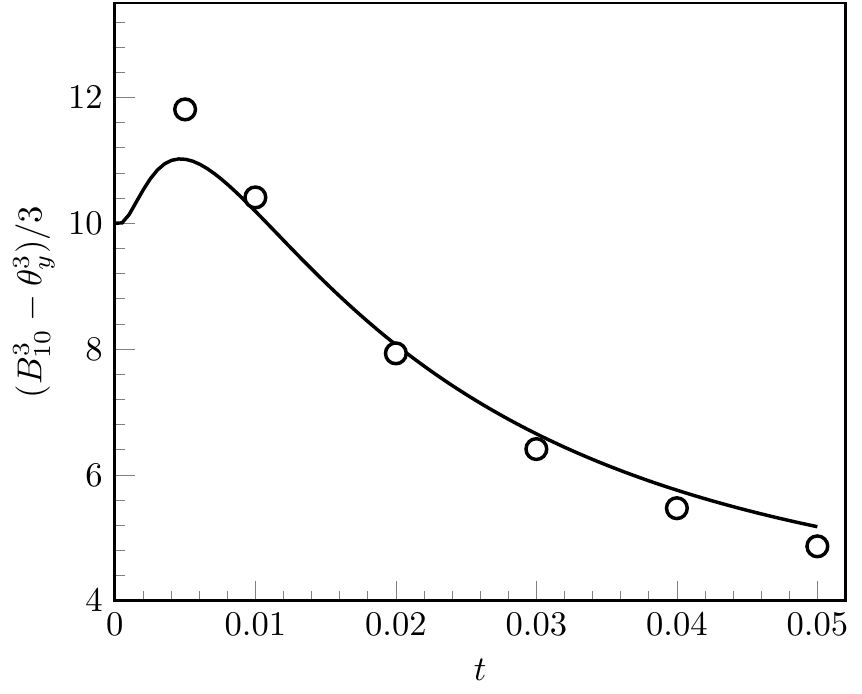}
\caption{Verification of the relation \eqref{anglespeed} between the contact angle
and the contact line velocity. The solid curve is 
$\left(B_{10}^3 -\theta_y^3\right)/3$ versus time, where $B_{10}(t)$ is computed
from the numerical solution of the leading order outer equation \eqref{H0eqn}.
The circles are the plot of $\dot{a}_1$ at different times, 
where the data for $\dot{a}_1$ are computed from the slope of the line jointing the
origin and the last data point $(\lambda=9\times 10^{-7})$
 in the lower panel of Figure \ref{fig:vel}. \label{fig:anglevel}}  
\end{figure}

Finally, a check of the angle-speed relation (\ref{anglespeed}) is given in Figure \ref{fig:anglevel}. The solid curve is the plot of $\left(B^3_{10}(t) -\theta_y^3\right)/3$, whereas the circles are the extracted values of $\dot{a}_1$ at the different times. These two sets of data agree well except for small time, $t$. We would expect that for a fixed value of $t$, the relation only holds in the limit $\lambda \to 0$. The error in the relation (\ref{anglespeed}) is due to our inability to resolve the contact line problem for sufficiently small values of slip.

\section{Breakdown as $t \to \infty$ and recovering the quasi-static limit} \label{hocking}

\noindent We notice that as $t \to \infty$, the above asymptotic analysis fails, since the expansion \eqref{aseries} becomes disordered once the correction to the contact line position, $a_1(t) = \Oh(1/\epsilon)$. In the double limit of $t \to \infty$ and $\lambda \to 0$, we have a distinguished limit which requires a re-scaling of time using $\tau = \epsilon t$. From (\ref{govX}), this gives
\begin{equation}
\epsilon \frac{\partial H}{\partial \tau} + 
\epsilon\dd{a}{\tau} \frac{\partial H}{\partial X} 
+ \frac{\partial}{\partial X}\left( H^2 \left( H + \lambda \right) 
\frac{\partial^3 H}{\partial X^3} \right) = 0.
\end{equation}

\noindent If we expand $H = H_0 + \epsilon H_1 + \ldots$, and the velocities, $da/d\tau = d a_1/d\tau + \epsilon d a_2/d\tau + \ldots$, then at leading order, we obtain a quasi-static solution, with $H_0 = \frac{3\kappa}{2a^3(\tau)} X[ 2a(\tau) - X]$, where $\kappa$ can be solved by applying conservation of mass and using the initial profile of the droplet, $\kappa = \int_0^a h(x,0) \ dx$. 

Notice that time dependence only enters into $H_0$ via the $a(\tau)$ term, and since all the subsequent orders depend solely on derivatives of the previous orders (with one term multiplying $da/d\tau$), then the profile shape only depends on time as a function of the droplet location. The classic quasi-static analysis then follows (\emph{c.f.} \cite{hocking_1981sliding} and other references in \S{\ref{literature}}). In this case, the full outer solution is given by 
\begin{multline}
H(X,a) = \frac{3\kappa}{2a^3} X\Bigl[2a - X\Bigr] + 
\epsilon \dd{a_1}{\tau} \biggl[ \frac{a^4}{9\kappa^2}\biggr] \biggl[ (2a - X)\log(2 a - X) \\ + X \log X - 2 a \log (2a) + \frac{3}{2a}X(2a - X)\biggr] + \cdots.
\end{multline}
%
Since the time dependent term $\partial H/\partial t$ only affects the outer analysis of the previous sections, then the inner and intermediate solutions, given by (\ref{H0inner}), (\ref{H1inner}), and (\ref{interm}), continue to be valid, and we have for the outer-to-inner and inner-to-outer limits,
\begin{subequations}
\begin{align}
H_{\text{out $\to$ in}} &= \biggl[\frac{3\kappa}{a^2}X + \ldots \biggr] + \epsilon \dd{a_1}{\tau} \frac{a^4}{9\kappa^2} \biggl[ X\log X + \Bigl\{2 - \log(2a)\Bigr\}X  + \ldots \biggr] + \cdots, \label{Hout2in} \\
\overline{H}_{\text{in $\to$ out}} &= \theta_y s + \epsilon \dd{a_1}{\tau}  
 \biggl[ \left(\frac{1}{\theta_y^2}\right)s \log s + \left( \frac{1}{\theta_y} + \frac{\log \theta_y}{\theta_y^2}\right) s\biggr] + \cdots. \label{Hin2out}
 \end{align}
\end{subequations}
If we denote $\theta_\text{app}$ as the leading order outer contact angle (the apparent contact angle), then we have from \eqref{Hout2in}, $\theta_\text{app} = 3\kappa/a^2$, which confirms that the apparent contact angle can be predicted once the contact line location is known. Using \eqref{c0c1}--\eqref{int2out}, \eqref{Hout2in}, and \eqref{Hin2out} allows the matching between inner and outer solutions through the intermediate layer giving
\begin{subequations}
\begin{gather}
\theta_\text{app}^3 - \theta_y^3 = 3\dd{a_1}{\tau}, \label{anglespeed2} \\
\dd{a_2}{\tau} = \dd{a_1}{\tau} \biggl[ -\beta \theta_y + \log\left( \frac{e^2}{2a\theta_y} \right)\biggr]. \label{da2dtau}
\end{gather}
\end{subequations}
which plays an analogous role to the two equations \eqref{anglespeed} and \eqref{anglespeed_more} for the $t = \Oh(1)$ problem. Using (\ref{anglespeed2}) and (\ref{da2dtau}), we then have a differential equation for the droplet location, accurate to two orders:
\begin{align}
\dd{a}{\tau} &\sim \dd{a_1}{\tau} \biggl[ 1 + \epsilon\biggl\{-\beta\theta_y + \log\left( \frac{e^2}{2a\theta_y} \right)\biggr\}\biggr] \notag \\
&= \frac{1}{3} \biggl[ \left(3\kappa/a^2\right)^3 - \theta_y^3 \biggr] 
\biggl[ 1 + \epsilon\biggl\{-\beta \theta_y + \log\left( \frac{e^2}{2a\theta_y} \right)\biggr\}\biggr], \label{ode_a}
\end{align}
This is analogous to the results of \cite{hocking_1983} using the alternative constitutive relationship \eqref{eq:CLcond}.

The principle result of this paper is now shown in Figure \ref{bigpic}. Here, the re-scaled velocity, $\dot{a}(t)/\epsilon$ is plotted as a function of time for the case of a spreading droplet with slip coefficient, $\lambda = 9 \times 10^{-7}$. The two time scales determining the dynamics are clearly visible (note the shaded region is only illustrative), we indeed confirm that the classical quasi-static approximation of \eqref{ode_a} is an excellent fit once time is appreciable. However, for $t = \Oh(1)$, the slip-free approximation of \eqref{anglespeed} will capture the correct dynamics. We note that because the second time scaling is only logarithmically large in the slip, $\lambda$, then for most practical values of the slip, the transition to the quasi-static regime occurs quite rapidly. However, as $\lambda \to 0$, we would indeed expect the transition point (\eg in Figure \ref{bigpic}) to move to infinity.

\section{Discussion} \label{discuss}

\noindent The difference between the two distinguished limits is well encapsulated in the two angle-speed relations (\ref{anglespeed}) and (\ref{anglespeed2}) which, though very similar in appearance, have completely  different interpretations:
\begin{subequations}
\begin{alignat}{3}
(i) \ & \lambda \to 0, \ t = t^* &\qquad \theta_\text{app}^3(t^*) - \theta_y^3 = 3\dd{a_1}{t}  \label{anglespeed_final1} \\
(ii) \ & \lambda \to 0, \ t = |\log \lambda| \tau &\qquad \theta_\text{app}^3[a(\tau)] - \theta_y^3 = 3 \dd{a_1}{\tau}  \label{anglespeed_class}
\end{alignat}
\end{subequations}

In the case of (i), where $t = \Oh(1)$, then the apparent contact angle is a \emph{known} function given by the solution of the leading-order no-slip equation (\ref{H0eqn}); thus, the leading-order slip velocity is also known, and in the limit $\lambda \to 0$, contact line slippage is a `regular' perturbative effect. By `regular', we mean that the contact line position tends to a constant, $a(t^*) \to a_0$ as $\lambda \to 0$. To leading order, one would say that the contact line is \emph{fixed}. Thus (\ref{anglespeed_final1}) provides a closed relation between the apparent angle and the first-order contact line speed once the $\lambda = 0$ equation has been solved. No microscopic properties are necessary in determining the contact line dynamics at this order. 

However, in the limit that $t \to \infty$, significant contact-line movement occurs, and the asymptotic relations used to derive (i) are invalid. Contact line movement can be brought-in by re-scaling time. Thus, in the case of (ii), where time is logarithmically large in the slip number, then the apparent angle is no longer a directly known value. It can only be computed once the droplet location, $a(\tau)$, is known, and this value must be found by solving an ordinary differential equation for the position, given by (\ref{ode_a}). Although the methodology which we have used to study the $t = \Oh(1)$ problem is very similar to the methodology as used in the classic quasi-static works of, for example, \cite{hocking_1983}, the principle motivation of our work is to highlight this idea of the non-uniformity within the time variable. 


Although the principal setting of our work was for the lubrication equations of thin film flow, the same ideas hold for slow viscous Stokes flow. The difficulty, however, is that even the simplest free-surface problems in Stokes flow are too unwieldy to solve, and so classical works on contact line dynamics in slow flow (\eg in \citealt{coxrg_1986}) have relied upon very general description of how the inner, intermediate, and outer asymptotics are performed. Moreover, we believe that other contact line models (\eg in Table \ref{tab:res}) will exhibit the same subtleties in their asymptotic analysis; the notion of a distinguished limit in time is a generic aspect that arises due to the separation of macroscopic and microscopic time scales. 

In a general problem, there may be multiple choices for the velocity scale and the resultant Capillary number. Consider a system that begins at $t = 0$ with an imposed (macroscopic) velocity scale of $U_\text{macro}$ (for example, this may correspond to forced flow through a channel with speed $U_\text{macro}$). In this case, this initial macroscopic velocity sets the Capillary number,
\begin{equation} \label{Camacro}
\Camac = \frac{\mu U_\text{macro}}{\sigma}. 	
\end{equation}
The leading-order contact line condition to impose on the outer flow is that the contact line is fixed. To an observer positioned away from the contact line, the contact line seems stationary, with surrounding bulk fluid moving at an $\Oh(1)$ velocity. This is emphasized by the analysis of the $t = \Oh(1)$ scaling of \S{\ref{outer}} for the case of lubrication theory, and where $U_\text{macro}$ corresponds to the initial relaxation speed of a droplet deposited far from its quasi-static state. 

However, at large times, the bulk fluid slows down from its initial relaxation velocity and is now moving at the same rate as the contact line. The macroscopic flow is now governed by a smaller Capillary number:
\begin{equation} \label{Cacl}
\Cacl = \epsilon \Camac,	
\end{equation}
where $\epsilon$ is the contact line velocity. Relative to this velocity scale, the inner limit of the outer velocity field must now take into account contact line movement. It can be seen by examining the slow flow equations and free-surface conditions that in the limit $\Ca_\text{cl} \to 0$, the fluid interface is flat to leading order. In essence, this justifies the assumptions found in the slow-flow contact line analysis of \cite{coxrg_1986} where the leading-order outer solution consisted of flow in a fixed wedge for small Capillary number flow.  

\subsection{Problems with patching between time-dependent and quasi-static regions}

\begin{figure} \centering 
\includegraphics{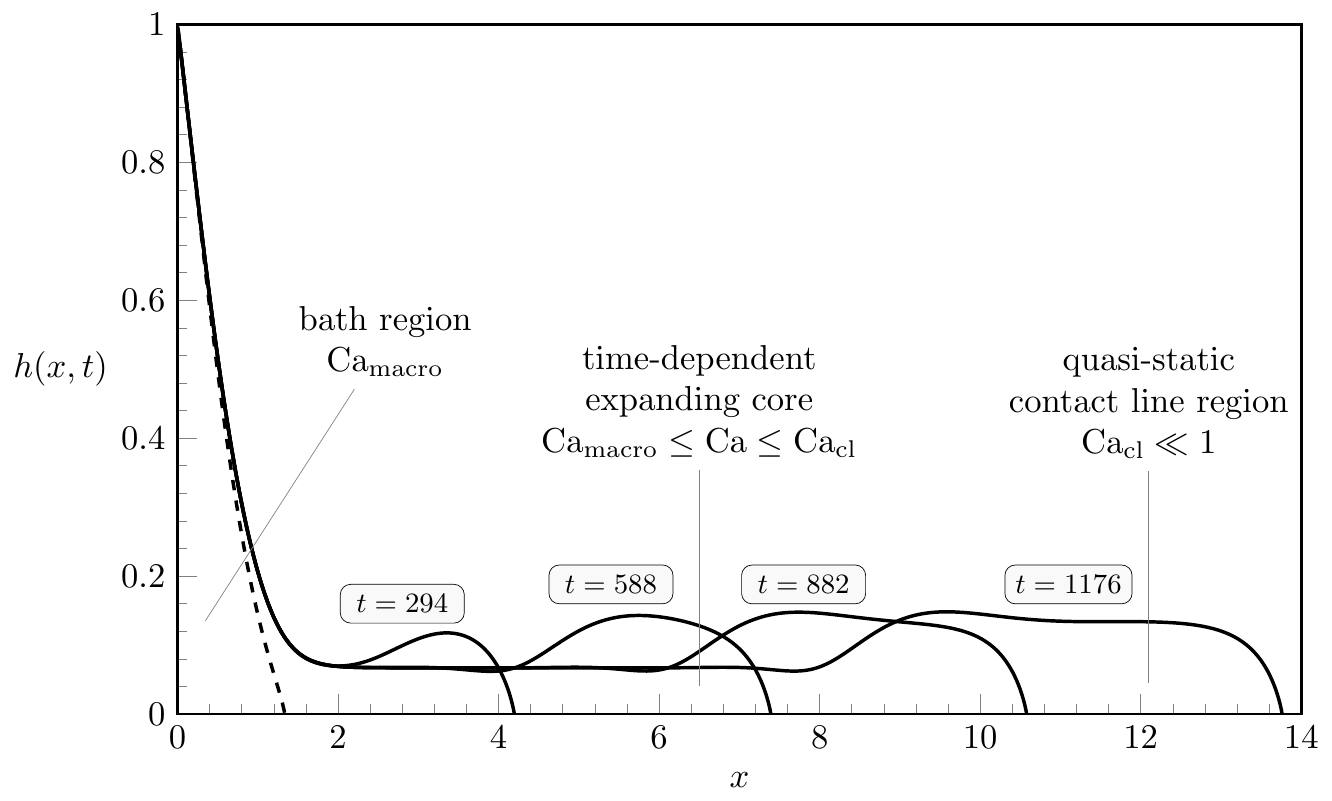}
\caption{Thin film profiles for a plate pulled from a bath, modeled by \eqref{pulledeqn} using $\Ca_\text{macro} = 0.017$ and $\lambda = 1.3 \times 10^{-5}$. The dashed line is the initial profile. \label{fig:pulled}}
\end{figure}

The above discussion highlights the difficulties of studying contact-line dynamics in situations where in the large time limit, the quasi-static flow near the contact line is not the entirety of the flow. Consider the situation of a plate pulled from a bath. Such a scenario is described by the non-dimensional thin film equation
\begin{equation} \label{pulledeqn}
\pd{h}{t} + \pd{}{x}\left[h^2\left(\frac{h}{3} +\lambda\right)\left(\pd{\kappa}{x} - 1\right) + \Ca_\text{macro} h \right] = 0,
\end{equation}
on the domain $x \in [0, R(t)]$, where we use the full nonlinear curvature $\kappa = h_{xx}/[1 + (h_x)^2]^{3/2}$. This classic dewetting problem has been studied by, for example, \cite{eggers_2004, eggers_2005existence}, \cite{eggers_2004characteristic}, and \cite{snoeijer_2006}. The point $x = 0$ corresponds to the matching solution to the bath, and we use the boundary conditions $h = 1$ and $h_x = -1$; on the right, at $x = R(t)$, we apply the contact line conditions \eqref{contcond}. In Figure \ref{fig:pulled}, we plot solutions at different times, corresponding to the capillary number, $\Ca_\text{macro} = 0.017$, and slip $\lambda = 1.3 \times 10^{-5}$. These values  have been chosen specifically to demonstrate the fascinating structure of the solution, and they have been used in \cite{snoeijer_2006}. 

In the limit that $t \to \infty$, it is seen that the bulk fluid near the bath tends to the `Landau-Levich solution' (\emph{c.f.} \citealt{wilson_1982}), where the plate is covered by a uniform film governed by the macroscopic capillary number $\Camac$. However, in a localized region near the contact line, the flow is increasingly quasi-static as $t \to \infty$, and the governing capillary number $\Cacl$ tends to zero as the slip is taken to zero. The size of this quasi-static region grows as time increases, and a contact line analysis would require matching the solution near the bath with solution near the contact line, through an intermediate time-dependent region whose length is \emph{a priori} unknown. Compare and contrast this with the situation of a spreading droplet, where in the limit $t \to \infty$, the leading order solution is globally solved by the quasi-static solution with constant curvature. 

The time-dependent drag-out problem has been studied by, for example, \cite{snoeijer_2006, snoeijer_2008}, and there, it was shown that the analysis is complicated further by the possibility of multiple solutions at large times. A similar system was studied in the work of \cite{benilov_2010}, where they demonstrated that for such problems, there exists an infinite number of zones, logarithmically spaced apart, where the fluid height oscillates between maximums and minimums. The key aspect of such problems is that, because the solution is only quasi-static near the very tip, an asymptotic analysis of the sort we have done here for $t = \Oh(1)$, is made difficult due to the required patching of multiple regions changing in time. Indeed, problems such as the case of gravity-driven draining down a vertical wall may not possess a well-defined limit as $\lambda \to 0$ and $t = \Oh(1)$, which is evident in the overturning profiles of \cite{moriarty_1991unsteady}. For such problems, not only is one required to contend with distinguished limits in time, as we have done in this paper, but also distinguished limits between the two (or more) capillary numbers. Such problems with more complicated global structure is the subject of ongoing investigation.

\mbox{}\par
\noindent \emph{Acknowledgements:} We gratefully thank the referees for their insightful comments and suggestions. We also thank Profs. Jens Eggers (Bristol), Howard A. Stone (Princeton), James M. Oliver (Oxford), and Andreas M\"{u}nch (Oxford) for many helpful discussions during the course of this work. The work of Ren was partially supported by Singapore A*STAR SERC PSF grant (project no. 1321202071)

\appendix
\section{Numerical Methods} \label{sec:num}

\noindent To solve the thin film equation \eqref{eq:heq} [and its reductions, such as \eqref{H0eqn}] on the time-dependent 
domain $[0,a(t)]$, where $a(t)$ is the moving contact line, we 
introduce the coordinate transformation: 
\begin{equation}
x(\xi,t) =a(t) f(\xi),
\end{equation}
where the map $f(\xi):[0,1]\rightarrow [0,1]$ is monotonic and $f(0)=0$, $f(1)=1$. The purpose of the map $f$ is to concentrate most of the grid points near the contact line. In this work, we use $f(\xi) =\tanh (\xi/\varepsilon)/\tanh (1/\varepsilon)$ where $\varepsilon = 0.2$.

In terms of the new variable, the thin film equation  becomes 
\begin{equation} \label{eq:heq1}
\frac{\partial h}{\partial t} - \frac{x_t}{x_\xi} \frac{\partial h}{\partial \xi}
+\frac{1}{x_\xi}\frac{\partial}{\partial \xi} \left(h^2 ( h+\lambda) 
\left(\alpha\frac{\partial h}{\partial\xi} 
+ \beta \frac{\partial^2 h}{\partial \xi^2 }
+\gamma\frac{\partial^3 h}{\partial\xi^3}\right)\right ) =0,
\end{equation}
where we have introduced
\begin{equation}
\alpha = -\frac{x_{\xi\xi\xi}}{x_\xi^4}+\frac{3 x_{\xi\xi}^2}{x_\xi^5},\quad
\beta = -\frac{3 x_{\xi\xi}}{x_\xi^4},\quad
\gamma = \frac{1}{x_\xi^3},
\end{equation}
and subscripts are used for partial derivatives.

Equation \eqref{eq:heq1} is solved on a uniform mesh covering the fixed domain 
$\xi\in [0,1]$ and  $t\in [0,T]$.
The solution is computed on the mid gird points 
$(\xi_{i+1/2}, t_n) = \left((i+1/2)\Delta \xi, n\Delta t\right)$,
where $\Delta \xi =1/N$ and $t_n = 1/M$ are the mesh steps in space and
time respectively. The numerical solution is denoted by $h_{i+1/2}^n$.

We use a semi-implicit scheme to evolve $h$ in time:
\begin{align} \label{eq:hn}
&\frac{h_{i+1/2}^{n+1}- h_{i+1/2}^n}{\Delta t} 
- \left(\frac{x_t}{x_\xi}\right)^n_{i+1/2}
\left(\frac{\partial h}{\partial\xi}\right)^{n+1}_{i+1/2}
+\left(\frac{1}{x_\xi}\right)^n_{i+1/2} 
\frac{R_{i+1}^{n+1} - R_{i}^{n+1}}{\Delta\xi} =0,
\end{align}
for $i=0,1,\cdots, N-1$. In the above equation, $R_i^n$ is the flux 
at the grid point $(\xi_i, t_n)$, which is given by
\begin{equation}
R^{n+1}_i = \left(h^2(h+\lambda)\right)_i^n 
\left(\alpha_i^n \left(\frac{\partial h}{\partial \xi}\right)_i^{n+1}
+ \beta_i^n \left(\frac{\partial^2 h}{\partial \xi^2}\right)_i^{n+1}
+ \gamma_i^n \left(\frac{\partial^3 h}{\partial \xi^3}\right)_i^{n+1}\right),
\end{equation}
for $i=1,2,\cdots,N-1$, and $R_0^{n+1} = R_N^{n+1} =0$.

The spatial derivatives are discretized using the standard finite differences:
\begin{subequations}
\begin{align}
\left(\frac{\partial h}{\partial\xi}\right)_{i+1/2}^{n+1}  &\approx \frac{1}{2 \Delta\xi}\left(h_{i+3/2}^{n+1} - 
   h_{i-1/2}^{n+1}\right),  \\
\left(\frac{\partial h}{\partial\xi}\right)_i^{n+1}  
  &\approx \frac{1}{\Delta\xi}\left(h_{i+1/2}^{n+1} - 
   h_{i-1/2}^{n+1}\right),  \\
\left(\frac{\partial^2 h}{\partial\xi^2}\right)_i^{n+1} 
  &\approx \frac{1}{2\Delta\xi^2}
  \left(h_{i+3/2}^{n+1} - h_{i-1/2}^{n+1} 
   -h_{i+1/2}^{n+1} +h_{i-3/2}^{n+1}\right), \\
\left(\frac{\partial^3 h}{\partial\xi^3}\right)_i^{n+1} 
&\approx \frac{1}{\Delta \xi^3}
\left( h_{i+3/2}^{n+1}-3h_{i+1/2}^{n+1} 
  +3 h_{i-1/2}^{n+1} - h_{i-3/2}^{n+1} \right).
\end{align}
\end{subequations}
Two ghost points are needed in order to evaluate the derivatives near the
boundary. They are defined using the boundary conditions \eqref{cond_0} and
\eqref{cond1}:
\begin{equation}
h_{-1/2}^{n+1} = h_{1/2}^{n+1}, \quad h_{N+1/2}^{n+1} = -h_{N-1/2}^{n+1}.
\end{equation}

In matrix form, the linear system in \eqref{eq:hn} has a banded structure, and it is easily solved using the $LU$ factorization to produce  $h_{i+1/2}^{n+1}$ for $i =0,1,\cdots, N-1$, the interface at the new time step. After the new interface is obtained, the contact line $a(t)$ is updated using the condition \eqref{eq:CLcond}.

\section{Complete Wetting} \label{perfectwetting}

\noindent The case of complete wetting, that is, $\theta_y = 0$ 
in \eqref{eq:CLcond}, requires a modification to the asymptotic analysis of \S\ref{outer}. If we assume again that the velocity is expanded into powers of $\epsilon$, then the degenerate boundary condition becomes $\partial H/\partial X = \Oh(\epsilon)$, and thus at first glance, the inner scaling of \S\ref{inner} would be such that the inner variables, $\overline{H}$ and $s$, satisfy $\partial \overline{H}/\partial s = \epsilon$; this would allow the wall-angle condition to be applied to the leading order inner solution. However, this is not the case, and one finds that such a scaling makes it impossible to perform the necessary matching between inner and outer solutions. 

In fact, the correct scaling for the inner region is such that the advective, capillary, and slip terms of (\ref{govX}) are all balanced at leading order. This requires $H = \lambda\overline{H}$ and $X = \lambda\epsilon^{-1/3} s$. Thus, for the case of complete wetting, the inner length scale is algebraically larger than that in the case of partial wetting. The inner solution is then expanded into the series, $\overline{H} = \overline{H}_0 + \epsilon \overline{H}_1 + \Oh(\epsilon^2)$, and the leading order problem satisfies
\begin{equation} \label{innerperfect}
\dot{a}_1 + \overline{H}_0 (\overline{H}_0 + 1) \frac{\partial^3 \overline{H}_0}{\partial s^3} = 0,
\end{equation}

\noindent with boundary conditions $\overline{H}_0(0) = \overline{H}'_0(0) = 0$. The third boundary condition is a matching condition. As it was shown by \cite{hocking_1992}, the outer limit of the leading order inner solution satisfies $\overline{H} \sim s [ 3\dot{a}_1 \log s + C ]^{1/3}$, where the value of $C$ is chosen to match the inner and outer solutions (through the intermediate layer). Because this involves the numerical solution of (\ref{innerperfect}), we have chosen to only present the details for the case of partial wetting; however, it should be clear that the main point of this paper (that of understanding the important of time re-scaling) continues to hold true, even for the case of complete wetting. 


\begin{thebibliography}{50}
\expandafter\ifx\csname natexlab\endcsname\relax\def\natexlab#1{#1}\fi

\bibitem[Anderson {\em et~al.\/}(1998)Anderson, McFadden \&
  Wheeler]{anderson_1998}
{\sc Anderson, D.~M., McFadden, G.~B. \& Wheeler, A.~A.} 1998 Diffuse-interface
  methods in fluid mechanics. {\em Ann. Rev. Fluid Mech.\/} {\bf 30}~(1),
  139--165.

\bibitem[Benilov {\em et~al.\/}(2010)Benilov, Chapman, McLeod, Ockendon \&
  Zubkov]{benilov_2010}
{\sc Benilov, E.~S., Chapman, S.~J., McLeod, J.~B., Ockendon, J.~R. \& Zubkov,
  V.~S.} 2010 On liquid films on an inclined plate. {\em J. Fluid Mech.\/} {\bf
  663}, 53--69.

\bibitem[Bertozzi \& Pugh(1994)]{bertozzi_1994}
{\sc Bertozzi, A.~L \& Pugh, M.} 1994 The lubrication approximation for thin
  viscous films: the moving contact line with a'porous media'cut-off of van der
  waals interactions. {\em Nonlinearity\/} {\bf 7}~(6), 1535.

\bibitem[Billingham(2008)]{billingham_2008}
{\sc Billingham, J.} 2008 Gravity-driven thin-film flow using a new contact
  line model. {\em IMA J. Applied Math.\/} {\bf 73}~(1), 4.

\bibitem[Blake(1993)]{blake_1993}
{\sc Blake, T.~D.} 1993 {\em Wettability\/}, {\em Surfactant Science Series\/},
  vol.~49, chap. Dynamic Contact Angles and Wetting Kinetics, p. 251. New York:
  Marcel Dekker, Inc.

\bibitem[Blake(2006)]{blake_2006}
{\sc Blake, T.~D.} 2006 The physics of moving wetting lines. {\em J. Colloid.
  Interface Sci.\/} {\bf 299}, 1--13.

\bibitem[Blake {\em et~al.\/}(1999)Blake, Bracke \& Shikhmurzaev]{blake_1999}
{\sc Blake, T.~D., Bracke, M. \& Shikhmurzaev, Y.~D.} 1999 Experimental
  evidence of nonlocal hydrodynamic influence on the dynamic contact angle.
  {\em Phys. Fluids\/} {\bf 11}, 1995--2007.

\bibitem[Blake {\em et~al.\/}(1997)Blake, Clarke, De~Coninck \&
  de~Ruijter]{blake_1997}
{\sc Blake, T.~D., Clarke, A., De~Coninck, J. \& de~Ruijter, M.~J.} 1997
  Contact angle relaxation during droplet spreading: comparison between
  molecular kinetic theory and molecular dynamics. {\em Langmuir\/} {\bf
  13}~(7), 2164--2166.

\bibitem[Blake \& De~Coninck(2002)]{blake_2002}
{\sc Blake, T.~D. \& De~Coninck, J.} 2002 The influence of solid--liquid
  interactions on dynamic wetting. {\em Adv. Colloid Interface Sci.\/} {\bf
  96}~(1), 21--36.

\bibitem[Blake \& Haynes(1969)]{blake_1969}
{\sc Blake, T.~D. \& Haynes, J.~M.} 1969 Kinetics of liquid/liquid
  displacement. {\em J. Colloid Interf. Sci.\/} {\bf 30}~(3), 421--423.

\bibitem[Cox(1986)]{coxrg_1986}
{\sc Cox, R.~G.} 1986 The dynamics of the spreading of liquids on a solid
  surface. {P}art 1. {V}iscous flow. {\em J. Fluid Mech.\/} {\bf 168},
  169--194.

\bibitem[De~Coninck \& Blake(2008)]{coninck_2008}
{\sc De~Coninck, J. \& Blake, T.~D.} 2008 Wetting and molecular dynamics
  simulations of simple liquids. {\em Annu. Rev. Mater. Res.\/} {\bf 38},
  1--22.

\bibitem[Dussan~V. \& Davis(1974)]{dussan_1974}
{\sc Dussan~V., E.~B. \& Davis, S.~H.} 1974 On the motion of a fluid-fluid
  interface along a solid surface. {\em J. Fluid Mech.\/} {\bf 65}, 71.

\bibitem[Eggers(2004{\natexlab{{\em a\/}}})]{eggers_2004}
{\sc Eggers, J.} 2004{\natexlab{{\em a\/}}} Hydrodynamic theory of forced
  dewetting. {\em Phys. Rev. Lett.\/} {\bf 93}~(9), 094502.

\bibitem[Eggers(2004{\natexlab{{\em b\/}}})]{eggers_2004toward}
{\sc Eggers, J.} 2004{\natexlab{{\em b\/}}} Toward a description of contact
  line motion at higher capillary numbers. {\em Phys. Fluids\/} {\bf 16}~(9),
  3491--3494.

\bibitem[Eggers(2005)]{eggers_2005existence}
{\sc Eggers, J.} 2005 Existence of receding and advancing contact lines. {\em
  Phys. Fluids\/} {\bf 17}~(8), 082106.

\bibitem[Eggers \& Stone(2004)]{eggers_2004characteristic}
{\sc Eggers, J. \& Stone, H.~A.} 2004 Characteristic lengths at moving contact
  lines for a perfectly wetting fluid: the influence of speed on the dynamic
  contact angle. {\em J. Fluid Mech.\/} {\bf 505}, 309--321.

\bibitem[Flitton \& King(2004)]{flitton_2004}
{\sc Flitton, J.~C. \& King, J.~R.} 2004 Surface-tension-driven dewetting of
  {N}ewtonian and power-law fluids. {\em J. Eng. Math.\/} {\bf 50}~(2-3),
  241--266.

\bibitem[Greenspan(1978)]{greenspan_1978}
{\sc Greenspan, H.~P.} 1978 On the motion of a small viscous droplet that wets
  a surface. {\em J. Fluid Mech.\/} {\bf 84}~(01), 125--143.

\bibitem[Haley \& Miksis(1991)]{haley_1991}
{\sc Haley, P.~J. \& Miksis, M.~J.} 1991 The effect of the contact line on
  droplet spreading. {\em J. Fluid Mech.\/} {\bf 223}~(57-81), 134.

\bibitem[Hocking(1981)]{hocking_1981sliding}
{\sc Hocking, L.~M.} 1981 Sliding and spreading of thin two-dimensional drops.
  {\em Q. J. Mech. Appl. Math.\/} {\bf 34}~(1), 37--55.

\bibitem[Hocking(1983)]{hocking_1983}
{\sc Hocking, L.~M.} 1983 The motion of a drop on a rigid surface. In {\em
  Proc. of the 2d Intern. Colloq. on Drops and Bubbles, Monterey.\/}, pp.
  315--321. JPL Publications.

\bibitem[Hocking(1992)]{hocking_1992}
{\sc Hocking, L.~M.} 1992 Rival contact-angle models and the spreading of
  drops. {\em J. Fluid Mech.\/} {\bf 239}~(1), 671--681.

\bibitem[Hocking \& Rivers(1982)]{hocking_1982}
{\sc Hocking, L.~M. \& Rivers, A.~D.} 1982 The spreading of a drop by capillary
  action. {\em J. Fluid Mech.\/} {\bf 121}~(1), 425--442.

\bibitem[Huh \& Scriven(1971)]{huh_1971}
{\sc Huh, C. \& Scriven, L.E.} 1971 Hydrodynamic model of steady movement of a
  solid/liquid/fluid contact line. {\em J. Colloid. Interface Sci.\/} {\bf 35},
  85--101.

\bibitem[Jacqmin(2000)]{jacqmin_2000}
{\sc Jacqmin, D.} 2000 Contact-line dynamics of a diffuse fluid interface. {\em
  J. Fluid Mech.\/} {\bf 402}~(1), 57--88.

\bibitem[King \& Bowen(2001)]{king_2001}
{\sc King, J.~R. \& Bowen, M.} 2001 Moving boundary problems and non-uniqueness
  for the thin film equation. {\em Eur. J. Appl. Math.\/} {\bf 12}~(03),
  321--356.

\bibitem[Kistler(1993)]{kistler_1993}
{\sc Kistler, S.F.} 1993 {\em Wettability\/}, {\em Surfactant Science
  Series\/}, vol.~49, chap. Hydrodynamics of Wetting, pp. 311--430. New York:
  Marcel Dekker, Inc.

\bibitem[Koplik {\em et~al.\/}(1988)Koplik, Banavar \& Willemsen]{koplik_1988}
{\sc Koplik, J., Banavar, J.~R. \& Willemsen, J.~F.} 1988 Molecular dynamics of
  poiseuille flow and moving contact lines. {\em Phys. Rev. Lett.\/} {\bf
  60}~(13), 1282--1285.

\bibitem[Koplik {\em et~al.\/}(1989)Koplik, Banavar \& Willemsen]{koplik_1989}
{\sc Koplik, J., Banavar, J.~R. \& Willemsen, J.~F.} 1989 Molecular dynamics of
  fluid flow at solid surfaces. {\em Phys. Fluids A\/} {\bf 1}, 781.

\bibitem[Lacey(1982)]{lacey_1982}
{\sc Lacey, A.~A.} 1982 The motion with slip of a thin viscous droplet over a
  solid surface. {\em Stud. Appl. Math\/} {\bf 67}~(3), 217--230.

\bibitem[Lauga {\em et~al.\/}(2007)Lauga, Brenner \& Stone]{lauga_2007}
{\sc Lauga, E., Brenner, M.~P. \& Stone, H.~A.} 2007 {\em Microfluidics: The
  no-slip boundary condition Handbook of Experimental Fluid Dynamics, C.
  Tropea, A. Yarin, JF Foss\/}, chap.~19, pp. 1219--1240. Springer, New York.

\bibitem[Moriarty \& Schwartz(1992)]{moriarty_1992}
{\sc Moriarty, J.~A. \& Schwartz, L.~W.} 1992 Effective slip in numerical
  calculations of moving-contact-line problems. {\em J. Eng. Math.\/} {\bf
  26}~(1), 81--86.

\bibitem[Moriarty {\em et~al.\/}(1991)Moriarty, Schwartz \&
  Tuck]{moriarty_1991unsteady}
{\sc Moriarty, J.~A., Schwartz, L.~W. \& Tuck, E.~O.} 1991 Unsteady spreading
  of thin liquid films with small surface tension. {\em Phys. Fluids\/} {\bf
  3}~(5), 733--742.

\bibitem[Pismen(2002)]{pismen_2002}
{\sc Pismen, L.~M.} 2002 Mesoscopic hydrodynamics of contact line motion. {\em
  Colloids and Surfaces A: Physicochemical and Engineering Aspects\/} {\bf
  206}~(1), 11--30.

\bibitem[Pomeau(2002)]{pomeau_2002}
{\sc Pomeau, Y.} 2002 Recent progress in the moving contact line problem: a
  review. {\em Comptes Rendus Mecanique\/} {\bf 330}~(3), 207--222.

\bibitem[Qian {\em et~al.\/}(2003)Qian, Wang \& Sheng]{Qian_2003}
{\sc Qian, T., Wang, X.-P. \& Sheng, P.} 2003 Molecular scale contact line
  hydrodynamics of immiscible flows. {\em Phys. Rev. E\/} {\bf 68}~(1), 016306.

\bibitem[Ren \& E(2007)]{ren_2007}
{\sc Ren, W. \& E, W.} 2007 Boundary conditions for the moving contact line
  problem. {\em Phys. Fluids\/} {\bf 19}~(2).

\bibitem[Ren {\em et~al.\/}(2010)Ren, Hu \& E]{ren_2010cont}
{\sc Ren, W., Hu, D. \& E, W.} 2010 Continuum models for the contact line
  problem. {\em Phys. Fluids\/} {\bf 22}~(10), 102103.

\bibitem[Shikhmurzaev(1997)]{shik_1997}
{\sc Shikhmurzaev, Y.~D.} 1997 Moving contact lines in liquid/liquid/solid
  systems. {\em J. Fluid Mech.\/} {\bf 334}~(1), 211--249.

\bibitem[Shikhmurzaev(2007)]{shik_book}
{\sc Shikhmurzaev, Y.~D.} 2007 {\em Capillary flows with forming interfaces\/}.
  Chapman and Hall/CRC.

\bibitem[Snoeijer {\em et~al.\/}(2006)Snoeijer, Delon, Fermigier \&
  Andreotti]{snoeijer_2006}
{\sc Snoeijer, J.~H., Delon, G., Fermigier, M. \& Andreotti, B.} 2006 Avoided
  critical behavior in dynamically forced wetting. {\em Phys. Rev. Lett.\/}
  {\bf 96}~(17), 174504.

\bibitem[Snoeijer {\em et~al.\/}(2008)Snoeijer, Ziegler, Andreotti, Fermigier
  \& Eggers]{snoeijer_2008}
{\sc Snoeijer, J.~H., Ziegler, J., Andreotti, B., Fermigier, M. \& Eggers, J.}
  2008 Thick films of viscous fluid coating a plate withdrawn from a liquid
  reservoir. {\em Phys. Rev. Lett.\/} {\bf 100}~(24), 244502.

\bibitem[Thompson \& Robbins(1989)]{thompson_1989}
{\sc Thompson, P.~A. \& Robbins, M.~O.} 1989 Simulations of contact-line
  motion: slip and the dynamic contact angle. {\em Phys. Rev. Lett.\/} {\bf
  63}~(7), 766--769.

\bibitem[Van~Dyke(1975)]{vandyke_book}
{\sc Van~Dyke, M.} 1975 {\em Perturbation Methods in Fluid Mechanics: By Milton
  Van Dyke. Annotated Ed\/}. Parabolic Press.

\bibitem[Velarde(2011)]{velarde_2011aa}
{\sc Velarde, M.~G.} 2011 Discussion and debate: Wetting and spreading science
  - quo vadis? {\em The European Physical Journal Special Topics\/} {\bf
  197}~(1), 1--148.

\bibitem[Voinov(1976)]{voinov_1976}
{\sc Voinov, O.~V.} 1976 Hydrodynamics of wetting. {\em Fluid Dynamics\/} {\bf
  11}~(5), 714--721.

\bibitem[Wilson {\em et~al.\/}(2006)Wilson, Summers, Shikhmurzaev, Clarke \&
  Blake]{wilson_2006}
{\sc Wilson, M.C.T., Summers, J.L., Shikhmurzaev, Y.D., Clarke, A. \& Blake,
  T.D.} 2006 Nonlocal hydrodynamic influence on the dynamic contact angle:
  {S}lip models versus experiment. {\em Phys. Rev. E\/} {\bf 73}, 041606.

\bibitem[Wilson(1982)]{wilson_1982}
{\sc Wilson, S. D.~R.} 1982 The drag-out problem in film coating theory. {\em
  J. Eng. Math.\/} {\bf 16}~(3), 209--221.

\bibitem[Yue {\em et~al.\/}(2010)Yue, Zhou \& Feng]{yue_2010}
{\sc Yue, P., Zhou, C. \& Feng, J.~J.} 2010 Sharp interface limit of the
  {Cahn--Hilliard} model for moving contact lines. {\em J. Fluid Mech.\/} {\bf
  645}, 279--294.

\end{thebibliography}

\providecommand{\noopsort}[1]{}

\end{document}